\documentclass[conference]{IEEEtran}
\IEEEoverridecommandlockouts
\usepackage{cite}
\usepackage{amsmath,amssymb,amsfonts}
\usepackage{algorithmic}
\usepackage{graphicx}
\usepackage{textcomp}
\usepackage{xcolor}
\usepackage{hyperref}
\def\BibTeX{{\rm B\kern-.05em{\sc i\kern-.025em b}\kern-.08em
    T\kern-.1667em\lower.7ex\hbox{E}\kern-.125emX}}
\begin{document}

\title{HyperPubSub: Blockchain based Publish/Subscribe}

\author{\IEEEauthorblockN{Gewu Bu}
\IEEEauthorblockA{\textit{LIP6} \\
\textit{Sorbonne University}\\
Paris, France \\
Gewu.Bu@lip6.fr}
\and
\IEEEauthorblockN{Thanh Son Lam Nguyen}
\IEEEauthorblockA{\textit{LIP6} \\
\textit{Sorbonne University}\\
Paris, France \\
Thanh-Son-Lam.Nguyen@lip6.fr}
\and
\IEEEauthorblockN{Maria Potop-Butucaru}
\IEEEauthorblockA{\textit{LIP6} \\
\textit{Sorbonne University}\\
Paris, France \\
maria.potop-butucaru@lip6.fr}
\and
\IEEEauthorblockN{Kim Loan Thai}
\IEEEauthorblockA{\textit{LIP6} \\
\textit{Sorbonne University}\\
Paris, France \\
kim.thai@lip6.fr}
}

\maketitle

\begin{abstract}
In this paper we describe the architecture and the implementation of a broker based publish/subscribe system where the broker role is played by  a private blockchain, Hyperledger Fabric. We show the effectiveness of our architecture by implementing and deploying a photo trading plateform. Interestingly, our architecture is generic enough to be adapted to any digital asset trading.
\end{abstract}


\section{Introduction} 
Publish/subscribe (e.g.  \cite{Birman:1987:EVS:37499.37515}) is a communication primitive designed for large scale dynamic networks due to the loosely coupled interaction between the publishers and subscribers. In this framework, publishers produce events and subscribers express their interests through subscriptions. Any event matching the subscription is delivered to the corresponding subscriber. The matching procedure is executed by brokers, which are also responsible for the event delivery. In this way, publishers and subscribers are completely desynchronized in time and space.  
Publish/subscribe systems design follow two main research lines: topic-based and content-based systems. In topic-based systems events published on a specific topic are forwarded to all clients subscribed to this topic. In content-based systems  subscribers specify their interests using filters. Many publish/subscribe systems are based on a fixed infrastructure of reliable brokers.  
The \emph{broker} in a publish/subscribe system solves the following problem: instead of communicating to each other directly, both publishers and subscribers will connect to the broker. The broker receives informations sent by publishers then it filters and transfers received information to corresponding subscribers. One of the filtering mode is \emph{topic based} filtering.  In this case, generated information may have different topic labels defined by publishers. Subscribers will chose topics they like to follow. The broker filters received information by topic and delivers it to suitable subscribers.

Even though publish/subscribe systems have been designed as an alternative communication primitive for large scale churn prone systems their design philosophy can go beyond the sending/receiving messages. That is, in recent years, more and more digital assets are tradable products. 
Publishers and  subscribers of these valuable digital assets can be therefore considered as producers and respectively potential customers 
on a e-market place that will trade using a broker based publish/subscribe architecture. 

It should be noted that classic publish/subscribe system  do not aim at  trading. Therefore, they cannot be easily adapted to varying industrial requirement including security and privacy of both clients and providers.
 That  is, broker based publish/subscribe systems make easy the coupling between publishers and subscribers during the information delivery however, they still suffer from many drawbacks, some of them listed below: 
\begin{itemize}
\item Delivery failure.
During the delivery, if some subscribers are not available for receiving data, they will lost the information. The broker has to decide whether and when it has to  re-transfer the information. That requires brokers to keep connected with subscribers  and learn their availability.
\item Delivery slowdown.
When a hot topic having a lot of subscribers is published, the broker has to transfer this information to all its subscribers. That will increase the traffic load on the network and the delivery delay for individual subscribers will increase due to the network capacity limitation.
\item Security risks.
Lacking of a complete secure design, the broker  becomes the critical single-failure point. Additional security protocols are needed to ensure the reliability of the system.
\end{itemize}

In order to circumvent some of the above mentioned drawbacks we propose a blockchain based publish/subscribe architecture where the role of the broker is played by a private blockchain Hyperledger Fabric. 

\section{Basics of Hyperledger Fabric}
Hyperledger Fabric \cite{cachin2016architecture, Androulaki:2018:HFD:3190508.3190538} is an open-source private blockchain framework started at Linux Foundation, designed for business purpose and now managed by Digital Asset and IBM. As a private blockchain, Hyperledger Fabric shares common property with traditional blockchains such as Bitcoin \cite{nakamoto2008bitcoin}: information stored on the chain is safe and traceable. Moreover, all interactions history  are permanent and cannot be changed unless an adversary owns more computation power than all honest participants. 

Hyperledger Fabric is composed of the following building blocks: 
\begin{itemize}
\item Membership Service Provider.
As a private chain, participants in the Hyperledger Fabric can have certain trust between each others. 
The Membership Service Provider ($MSP$) module defines relations between participants and gives them different access rights. Only the nodes registered via $MSP$ can connect to Hyperledger Fabric. MSP provides a basic security indentification for the system.

\item Channel.
Instead of having one single blockchain, Hyperledger Fabric maintains simultaneously many blockchains, different types of information stored in different blockchains are independent. Participants can join into one or multi channels according to the business logic design. So that they can access to the blockchain corresponding to the channels they belong to. 

\item Chaincode.
For each channel, many Chaincodes (also called smart contracts) can be installed. Chaincodes are programs that can be invoked by all participants in the same channel to interact with the information stored into the channel. Note that interactions such as:  creation, modification or delete the information will be recorded in blockchain of the channel.

\item Ordering Service.
Ordering Service is a  consensus  module.  It connects to all the channels.  Ordering Service receives incoming invocations belonging to the same channel and order them into blocks. Then it diffuses these blocks to the participants of the channel, so that they can append the block into their local chain and update stored information according to the invoked chaincode.
\end{itemize}

\figurename~ \ref{fig:phases} describes how a block can be added into a channel in Hyperledger Fabric with three phases: Execution-Ordering-Validating: 

\begin{figure}[!hbt]
\centering
\includegraphics[width=\linewidth]{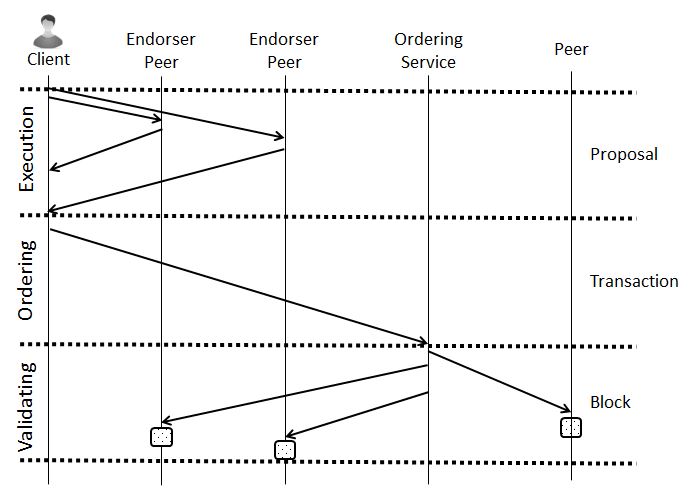}
\caption{Three phases architecture of Hyperledger Fabric.}
\label{fig:phases}
\end{figure}

\emph{Execution phase.} When invoking a chaincode of a channel, a user (called client) having access to this channel can send the  proposal to the endorser peers (the network participants who have the chaincode installed) and waits for their reply. When receiving the proposal, endorsers execute individually the program according to the invoked chaincode. Endorser peers  sign this endorsement with the result, send it back to the client as a proposal response. When the client receives enough responses, it creates an official invoke requirement (called transaction) that assembles these responses for the next phase.

\emph{Ordering phase.} A client sends its transaction to the ordering service after finishing the execution phase. The ordering service orders all submitted transactions per channel and combines these transactions into blocks.

\emph{Validating phase.} The ordering service broadcasts blocks to all the peers in a channel. Then each peer will verify each transaction in the block. Transactions failing the verification will be marked as invalid. After the checking, each peer appends the whole block (including invalid transactions) to its local blockchain. Note that invalid transaction can be distinguished, and will be considered as not existent, so that invalid invoke requirement will not affect the stored information.

Secure membership management, independent multi-channel structure and flexible chaincodes make Hyperledger Fabric become a powerful business logic oriented framework.

\section{HyperPubSub Design}
In this section we propose a new broker based Publish/Subscribe system, HyperPubSub,  designed on top of Hyperledger Fabric. HyperPubSub is specifically design to support trading of digital assets. In order to validate our architecture we designed a prototype of a online photo trading  plateform.

In $HyperPubSub$, publishers and subscribers are two types of clients in Hyperledger Fabric framework: one generates digital products, the other express their interest to consume them. The core of Hyperledger Fabric will be the broker in charge to  connect publishers and subscribers matching to the same topic by joining them into the corresponding channel. By combining chaincode and multi-blockchain structure, various topics can be managed safety and independently. In this way, the \emph{Delivery failure} and the \emph{Delivery slowdown} problem in classic Publish/Subscribe system can be avoided: subscribers don't need to receive the information passively. That is, they can query the information by invoking the chaincode proactively and individually.

When subscribers decide to buy a digital product published by publishers, the online trading can be done safely in trading dedicated channels, isolated from the other channels. All the trading history and details are stored into a  blockchain, so no one can cheat during the trading.

\subsection{Use Case: an Online Photo Trading Platform}

\begin{figure}[!hbt]
\centering
\includegraphics[width=\linewidth]{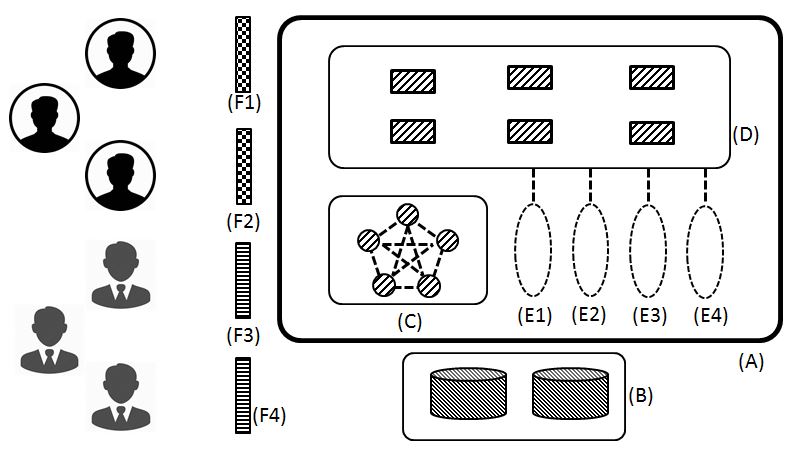}
\caption{Architecture of our trading platform.}
\label{fig2}
\end{figure}
In the following we instantiate $HyperPubSub$ in order to respond to a practical use case\footnote{This usecase was proposed to us by Andromak start-up. http://www.andromak.com}: an online photo trading platform that will bring together professional photographers and press agencies.  
Photographers (publishers in our architecture) upload their photos and set a price for their products; customers (subscribers in $HyperPubSub$) can review photos according to different categories (topics) and can choose to buy those they wish. This platform makes use of blockchain technology in order to guarantee that the metadata of a photo cannot be altered, in the same time it ensures the security of the online trading and the integrity of the photographers and customers identities.

In 
\figurename~\ref{fig2} we represent our architecture. The network topology is transparent to the clients. We have several gateways (F1-F2-F3-F4) that share the connections with the clients. Two database servers (B) are designed to store the photographs, one master and one slave for backup. The information of the clients, their photos, transactions, trades, etc. are managed by Hyperledger Fabric network (A). This blockchain network contains orderer nodes (C) which run Raft consensus \cite{ongaro2013search}. In our prototype we consider only one organization with six endorser peers nodes (D) and four channels (from E1 to E4). Each channel contains and runs one chaincode. There are four  independent blockchains: E1 channel stores the information of the clients; E2 channel contains the meta information of all photos; E3 channel is dedicated to all trades from customers to photographers; E4 channel stores extra information that facilitates the administration of this environment.

Some basic behaviors of two kinds of clients are defined and programmed. Each of them can create a new account and login.
A photographer can upload a photo from his device to data base (B), set this photo with three different prices for three different copyrights. This photo can be set in multiple categories (topics) (e.g. nature, sport, human, animal).

A customer can see all of the photos and additional information concerning the photographers so that the customer can choose one photo and buy it using  the HyperPubsub coins  (the crypto money in our system). A customer must posse an amount of coins superior  or equal to the price value of the photo he wants to trade. Therefore, we created an application for coins administration. The idea is that, a customer will use fiat to transfer money to the owner of the photo. After receiving the money, the owner of the platform use this application to provide a  customer with coins.   Here below is the description of a process to buy a photo:

\begin{itemize}
\item A customer uses  the application to login. This application will then connect to the gateway to transfer this customer's login information. Gateway then query to the our blockchain system to verify the user. In case the verification succeeds, the gateway queries the list of the photos from the blockchain system and sends the result to the customer.

\item After logging in, the customer can choose to show a specific photo category and then can chose the photo to buy. After validation, the application will connect and send a query to the gateway. The gateway then connect to channel E1, E3, E4 and sends corresponding chaincode  queries. After this step, the walled of the customer will be decreased and atomically the wallet of the photographer who owns that photo will be increased.

\item Data base server will receive the request for downloading the original photo. It will find that photo in its data base then send it back to the gateway such that the customer can download it.
\end{itemize}

\subsection{Implementation Environment for our prototype.}
 After having the architecture design above, we begin to implement in a real environment. 
 We implemented a prototype where some parts of the above described architecture have been omitted. 
 We have only one gateway for all photographers and customers and one database. We deployed our prototype  in two cloud instances, one in Germany (Heidelberg University) and one is in France (Onelab LIP6) using 2 vCPU (2.6GHz, 4GB of RAM, runs Ubuntu 16.04.5 LTS in each instance. Our Hyperledger Fabric blockchain network is 1.4.1 version with RAFT consensus. The application for client as well as database (file server) are programmed in java, and we NodeJs  for the gateway.

\section{Conclusion} 
In this paper we describe a broker based architecture for a publish/subscribe system where the broker role is played by Hyperledger Fabric. 
We instantiate our architecture for an online trading photo plateform and  implemented a prototype. Our prototype is currently deployed on two private clouds running in France and Germany. A demonstration video can be found here: \href{https://youtu.be/RFLmsSRmyMs}{https://youtu.be/RFLmsSRmyMs}.

\bibliographystyle{IEEEtran}
\bibliography{IEEEexample}

\end{document}